\documentclass[aps]{revtex4}

\usepackage{amsmath}
\usepackage{amssymb}
\usepackage[dvips]{graphicx}

\begin{document}

\title{Stability of Closed Timelike Geodesics in different Spacetimes.}

 \author{Val\'eria M. Rosa\footnote{e-mail: vmrosa@ufv.br}
  }

 \affiliation{
Departamento de Matem\'atica, Universidade Federal de Vi\c{c}osa, 
36570-000 Vi\c{c}osa, M.G., Brazil
}

\author{ Patricio  S. Letelier\footnote{e-mail: letelier@ime.unicamp.br}
 }

\affiliation{
 Departamento de Matem\'atica Aplicada-IMECC,
Universidade Estadual de Campinas,
13081-970 Campinas,  S.P., Brazil}                                                       
    
\begin{abstract}
 The linear stability of closed timelike geodesics (CTGs) is analyzed  in
 two spacetimes with cylindrical sources,  an infinite
 rotating dust cylinder, and  a cylindrical cloud
 of static cosmic strings with a central  spinning  string. 
We also study the existence and linear stability
 of closed timelike curves in spacetimes that share some common features with
 the G\"odel universe  (G\"odel-type spacetimes). In this case  the existence of CTGs
 depends on the   `background' metric.
The  CTGs  in a subclass of inhomogeneous stationary cosmological solutions of
 the Einstein-Maxwell equations  with topology  $ S^3\times \mathbb R$ are also examined. 

\end{abstract}
\maketitle

\section{Introduction}

The existence of closed timelike curves (CTCs) presents a clear
 violation of causality. In some cases these CTCs can be disregarded
 because to have them one ought to have an external force acting along
 the whole CTC, process that will consume a great amount of
 energy. The energy needed to travel a CTC in G\"odel universe is
 computed in \cite{pfarr}.  For geodesics this is not the case since
 the external force is null, therefore the considerations of energy
 does not apply in this case and we have a bigger problem of breakdown
 of causality.

To the best of our knowledge we have four solutions to the Einstein
 equations that contain CTGs. One of them was given by Bonnor and
 Steadman~\cite{bonnor} that studied the existence of CTGs in a
 spacetime with two spinning particles each one with magnetic moment
 equal to angular momentum and mass equal to charge (Perjeons), in
 particular, they present a explicit CTG. We found that this
 particular CTG is not stable, but there exist many other that are
 stable~\cite{rosalet1}. Soares~\cite{soares} found a class of cosmological
 models, solutions of Einstein-Maxwell equations, with a subclass
 where the timelike paths of the matter are closed. For these models
 the existence of CTGs is demonstrated and explicit examples are
 given. Steadman~\cite{steadman} described the existence of CTGs in a
 vacuum exterior of the van Stockum solution for an infinite rotating
 dust cylinder. For this solution explicit examples of CTCs and CTGs
 are shown. And in~\cite{gron} it is found CTGs in a spacetime
 associated to a cylindrical  cloud of static strings with negative mass density  with a central  spinning  string.

The possibility that a spacetime
associated to a realistic model of matter may contain CTCs  and, in particular, CTGs 
leads us to
ask how permanent is the existence of these curves. Perhaps, one may
rule out the CTCs by simple considerations about their linear
stability. Otherwise, if these curves are stable under linear
perturbations the conceptual problem associated  to their existence 
is enhanced. Even though the matter content of the solutions listed before are far from realistic we shall consider the study the stability of these curves 
in order to see the possibility to rule them out only by dynamical
 considerations. These considerations lead us to study the stability of CTCs in G\"odel 
universe \cite{vallet}. 

In the present work, besides the study of the stability under linear
perturbation of CTGs in the spacetimes described above, we shall also consider
the existence and stability under linear perturbation of CTCs in the two examples of G\"odel-type metrics given in
\cite{gurses}, see also \cite{gleiser}. One of them has only CTCs and the other has  CTGs
depending on the choice of the parameters. All the cases analyzed are
stationary and have axial symmetry. 

It is interesting to note that
these spacetimes are not counter examples of the Chronology Protection
Conjecture~\cite{cpc} that  essentially says that the laws of the
physics do not allow the appearance of closed timelike curves. The spacetimes that we shall considered are given  stationary spacetimes. A valid dynamic to built them is not known.

In  Section 2 we present the general equations that will used to study linear stability. In  Sections 3 and 4 we analyze the cases of a dust
cylinder and a cylinder of
cosmic strings with  a central spinning string, respectively. In  Section 5 we study two cases of G\"odel-type
metrics, one with flat background and the other with a conformally flat background. In
 Section 6 we consider  the two explicit examples of CTGs given
in~\cite{soares}. And finally, in  Section 7,  we discuss and summarize ours results.

\section{Stability of CTCs and CTGs}
As we mentioned before the stability of CTCs, for the G\"odel cosmological  model are studied in \cite{vallet}. Stability of geodesics  are studied in \cite{shirokov} for particles moving around a black hole. Also in \cite{semerak} and \cite{letlord} considered the stability of geodesic moving 
on accretion disks and other structures.

Excepting the Soares CTGs all the others  closed timelike geodesics that we shall study  belong to  spacetimes with
metrics,  $ds^2=g_{\mu\nu}dx^{\mu}dx^{\nu},$ where
$ x^{\mu}=[t,r,\varphi,z]$. In this case  all the curves  have the same  parametric form,
\begin{equation} 
t=t_*,\hspace{1cm} r=r_*, \hspace{1cm} \varphi \in [0,2\pi], \hspace{1cm}
z=z_* ,
\label{CTC}
\end{equation}
where $t_*$, $z_*$ and $r_*$ are constants. The condition for these curves to
 be timelike is  $\frac{dx^\mu}{d\varphi}\frac{dx_\mu}{d\varphi}>0,$ in 
other words,  $g_{\varphi\varphi} > 0$.

A generic CTC $\gamma$ satisfies the system of equations,
\begin{equation}
\ddot{x}^\mu+\Gamma^{\mu}_{\alpha \beta}\dot{x}^\alpha\dot{x}^\beta  = F^\mu(x),
\label{CTCsystem}
\end{equation}
where the overdot indicates derivation with respect to $s$,
$\Gamma^{\mu}_{\alpha \beta}$ are the Christoffel symbols and $F^\mu$
is a specific external force $(a^\mu=F^\mu)$. We have a closed
timelike geodesic when $a^\mu=0, \, \mu=0,1,2,3$.

To analyze the linear stability of a CTG $\gamma$ we consider  a small perturbation ${\bf \xi}$. The perturbed curve, $\tilde{\gamma}$, has the form  $\tilde{x}^{\mu}=x^{\mu}+\xi^{\mu}$.
From equations (\ref{CTCsystem}) one finds \cite{vallet} that the system of
differential equations satisfied by the perturbation ${\bf \xi}$
is,
\begin{equation}
\frac{d^2\xi^{\alpha}}{ds^2}+2\Gamma^{\alpha}_{\beta
\mu}\dot{x}^{\mu}\frac{d\xi^{\beta}}{ds}+\Gamma^{\alpha}_{\beta
\mu,\lambda}\dot{x}^{\beta}\dot{x}^{\mu}\xi^{\lambda}=F^{\alpha}_{,\lambda}\xi^{\lambda}\,\, ,
\label{Pertsystem}
\end{equation}
where $f_{,\lambda}\equiv\partial f/\partial x^{\lambda}$. The coefficients
of the metrics studied do not depend on $\varphi$. Therefore the
coefficients of the linear system (\ref{Pertsystem}) are all constants
and the analysis of stability in this case  reduces to the solution of a linear system of equations with constant coefficients. Also in all cases we will have conservation of the angular momentum that for the studied CTGs will
imply that the angular velocity $\dot{\varphi}$ will be a positive constant.

Excepting Soares CTGs, all the other are circles  in the  $(r,\varphi)$-plane. We will have a CTG
when   the function $s(r,z_*)=2\pi\sqrt{g_{\varphi\varphi}(r,z_*)}$  presents  a local maximum  at some value of $r$, say $r=\bar{r}$. It is important to stress that to
have a reasonable CTG it is necessary that $\bar{r} \ne 0$ and
$g_{\varphi\varphi}(\bar{r},z_*)>0$. Moreover, a way to obtain 
CTGs in a spacetime that only has CTCs is to deform this last spacetime,
for example,  adding matter in such way that the function
$g_{\varphi\varphi}(r,z_*)$ of the deformed spacetime be a function with a local maximum
at $r=\bar{r}$ with $\bar{r} \ne 0$ and
$g_{\varphi\varphi}(\bar{r},z_*)>0$.

\section{Van Stockum solution}

Steadman~\cite{steadman} described the behavior of CTGs in the
 exterior of the van Stockum solution for an infinite rotating dust
 cylinder. The metric is expressed in Weyl-Papapetrou coordinates as,
\begin{equation}
ds^2=Fdt^2-H(dr^2+dz^2)-Ld\varphi^2-2Md\varphi dt.
\label{vsmetric}
\end{equation}
The metric coefficients  in the  interior of the cylinder  are,
\begin{equation}
H=e^{-a^2r^2},\,L=r^2(1-a^2r^2),\,\rho=4a^2e^{a^2r^2},\,M=ar^2,\,F=1,
\end{equation}
where $a$ is the angular velocity of the cylinder and $\rho$  the
matter density.

In order to have no superluminal matter it is required that radius of
the cylinder be less than $1/a$, i.e., at the boundary  $r=R<1/a$ . For the  closed curve
$\gamma$ given in (\ref{CTC})   this
condition does not allow CTCs  inside the cylinder.

Van Stockum  found a procedure which generates an exterior
solution for all $aR>0$. He divided this solution in three
possibilities, depends on the value of $aR$. We have CTCs when
$aR>1/2$, in this case,  the exterior solution is
\begin{eqnarray}
&& H=e^{-a^2R^2}(r/R)^{-2a^2R^2}, \nonumber \\
&& L=\frac{Rr\sin(3\beta+\ln(r/R)\tan\beta)}{2\sin2\beta \cos \beta},\nonumber \\
&& M=\frac{r\sin(\beta+\ln(r/R)\tan\beta)}{\sin2\beta},\nonumber \\
&& F=\frac{r\sin(\beta-\ln(r/R)\tan\beta)}{R\sin\beta}, 
\label{metricfunctions}
\end{eqnarray}
with
\begin{equation}
\tan\beta=\sqrt{4a^2R^2-1},\;\; \frac{1}{2}<aR<1.\label{tanbeta}
\end{equation}
In the definition of $\tan\beta$, we take the positive square root and
the principal value of $\beta$.  With these restrictions, it is possible \cite{steadman} to
find closed timelike geodesics in this exterior solution.

For the exterior metric (\ref{metricfunctions}), the curve $\gamma$ is
timelike when $g_{\varphi\varphi}=-L>0$ and this occurs when $r_*$
belongs to open interval $R_k$, where
\begin{equation}
R_k=\Big(R\,\exp\Big[\frac{(2k-1)\pi-3\beta}{\tan\beta}\Big],\,\,R\,\exp\Big[\frac{2k\pi-3\beta}{\tan\beta}\Big] \Big),\,\,\,k\in\cal{Z}.
\end{equation}
The four-acceleration of $\gamma$ has only one non zero component,
\begin{equation}
a^r=\frac{e^{a^2R^2}(r_*/R)^{2a^2R^2}\sin(4\beta +\ln(r_*/R)\tan\beta)}{2r_*\cos\beta\,\sin(3\beta +\ln(r_*/R)\tan\beta)}.
\end{equation}
 The radial coordinates of geodesics are the
solutions of $a^r(r_*)=0$. There are an infinite number of  solutions and
those occurring in the regions $R_k$ are,
\begin{equation}
r_*=r_k=R\,e^{2(k\pi-2\beta)\cot\beta}, \, k=1,\,2,\,3,...
\end{equation}
The radial coordinates of the CTGs coincide with the local maximun
of $g_{\varphi\varphi}$.

For the above mentioned closed timelike geodesics the
system (\ref{Pertsystem}) reduces to
\begin{eqnarray}
&&\ddot{\xi}^0+k_1\dot{\xi}^1=0, \nonumber \\
&&\ddot{\xi}^1+k_2\dot{\xi}^0+k_3\xi^1=0,\nonumber \\
&&\ddot{\xi}^2+k_4\dot{\xi}^1=0,\nonumber \\ 
&&\ddot{\xi}^3=0,
\label{PertSystem1}
\end{eqnarray}
where 
\begin{equation}
k_1=2\Gamma^0_{21}\dot{\varphi},\, k_2=2\Gamma^1_{20}\dot{\varphi},\,
 k_3=\Gamma^1_{22,1}\dot{\varphi}^2, \, k_4=2\Gamma^2_{21}\dot{\varphi}.
\end{equation}

The solution of (\ref{PertSystem1})  is
\begin{equation}
\begin{array}{l}
\xi^0=-k_1(c_3\sin(\omega s+c_4)/\omega+\lambda s)+c_1\,s+c_5,\\ 
\xi^1=c_3\cos(\omega s+c_4)+\lambda, \\
\xi^2=-k_4(c_3\sin(\omega s+c_4)/\omega+\lambda s)+c_2\,s+c_6,\\ 
\xi^3=c_7\,s+c_8, \label{sol1}
\end{array}
\end{equation}
where $c_i,\;i=1,\dots,8$ are integration constants,
\begin{eqnarray}
\omega &=&\sqrt{k_3-k_1k_2}\\
 &=&\Big[\Big(\frac{r_k}{R}\Big)^{2a^2R^2-1}  \frac{e^{a^2R^2}}{8\cos^4 \beta}\dot{\varphi}^2\Big]^{1/2},
\end{eqnarray}
and $\lambda = -k_2c_1/\omega^2$. 

Since  $\omega>0$ is real the solution  (\ref{sol1}) shows  the typical behavior  for    stability, we have vibrational modes untangled with translational ones that can be 
eliminated by a suitable choice of the initial conditions.

\section{Cloud of cosmic strings }

Now we shall consider the spacetime associated to an infinite  cylinder   formed by 
a  cloud of parallel static strings with   a central  spinning cosmic
 string~\cite{gron}. The clouds of cosmic strings were introduced in \cite{let1}, see  also \cite{let2},  and for spinning strings see \cite{gallet,tl}.  
The  spacetime related to the interior of the cylinder of finite radius, $r=R$, is matched continuously to the external metric that is taken as representing a rotating cosmic string. 

 The line element
inside cylinder is 
\begin{equation}
ds^2=(dt-k\,d\varphi)^2-[D^2(r)d\varphi^2+dr^2+dz^2].
\label{gj}
\end{equation}

 We have  closed timelike curves when
$k^2-D(r_*)>0$. The nonzero component of the four-acceleration of this
curve is given by $a^r=D'(r)$.  Therefore, when $D'(r_*)=0$ the curve   $\gamma$   is a geodesic.

The nonzero contravariant components of the energy-momentum tensor in
the cloud  are $T^{tt}=\rho$ and $T^{zz}=p=-\rho$. The Einstein
equations in this case reduce to the single equation,
\begin{equation}
\dfrac{D''}{D}=-\rho.
\label{signD}
\end{equation}

 The sign  analysis  of $D''(r)$ and $D(r)$ shows that if there
exist CTGs and  the condition $D''(r)D(r)<0$  holds, then $D(r)$ changes sign at least once. Therefore, there are values of $r=\bar{r}$ where $D(\bar{r})=0, $ i.e., the metric is
degenerate. In order to obtain a connected spacetime it is assumed
that the mass density of the cloud  is negative. Hence $D''(r)D(r)>0$ for
$r\le R$.

For the metric (\ref{gj}), the system 
(\ref{Pertsystem}) is
\begin{eqnarray}
&&\ddot{\xi}^0=0,\nonumber \\
&&\ddot{\xi}^1-D''(r_*)D(r_*)\dot{\varphi}^2\xi^1=0 \nonumber \\
&&\ddot{\xi}^2=0,\,\,\ddot{\xi}^3=0.
\label{PertSystem2}
\end{eqnarray}
We have a solution with periodic modes when $D''(r_*)D(r_*)<0$. But it
was assumed that $D''(r)D(r)>0$ for $r\le R$, then the CTGs inside
the cylinder are not stable.

The line element  outside the cylinder is assumed to be the one of a spinning string,
\begin{equation}
ds^2=(dt-8\pi J d\varphi)^2-(1-8\pi\lambda)^2r^2d\varphi^2-(dr^2+dz^2),
\end{equation}
where $\lambda$ is the  string linear density and  $J$ the spin angular momentum per length unit.
 There exist CTCs outside cloud  when
$r_*<8\pi|J|/|1-8\pi\lambda|$, but, we do not have CTGs. This can be proved by  analyzing the  existence of  CTCs with maximum length outside of the cylinder. The
function that gives the length of these CTCs is $s(r)=2\pi
[(8\pi J)^2 -(1-8\pi\lambda)^2r^2]^{1/2}$ that has maximum
point only at  $r=0$.
Moreover, the length of CTCs inside  the cylinder is given by
$s(r)=2\pi \sqrt{k^2-D^2(r)}$ and  in order to have a CTG we need a
function $D(r)$ with a local maximum at $\bar{r} \ne 0$ such that
$k^2-D^2(\bar{r})>0$. This occurs for the two examples given
in~\cite{gron}. They are (i) $D(r)=\beta \cosh[(r-R)/r_0+\alpha]$ and
(ii) $D(r)=c[(r-a)^2+b^2]$, where $\alpha$, $\beta$, $a$, $b$, $c$, and
$r_0$ are constants.

\section{The G\"odel-type cases}

A G\"odel-type (GT) metric $g_{\mu\nu}$, as defined in~\cite{gurses}, is a
$D$-dimensional metric of the form
\begin{equation}
g_{\mu\nu}=u_{\mu}u_{\nu}-h_{\mu\nu},
\label{general_gtm}
\end{equation}
where the `background' $h_{\mu\nu}$ is the metric of
a $(D-1)$-dimensional spacetime perpendicular to the   timelike unit
vector $u^{\mu}$. Further more we assume that $h_{\mu\nu}$ and $u_{\mu}$ are
independent of the fixed special coordinate $x^k$ with $0 \le k \le
D-1$ and, moreover, that $h_{k\mu}=0$.

We shall  consider the special  cases with  $D=4$ (four dimensional spacetime) and constant $u_k$ . Also we  assume, without losing  generality, that the special fixed coordinate $x^k$ is $x^0\equiv t$, then  $h_{0\mu}=0$. We also do $u_0=1$.

The G\"odel-type metric (\ref{general_gtm}) solves the
Einstein-Maxwell dust field equations in four dimensions provides the
flat three-dimensional Euclidean source-free Maxwell equations
\begin{equation}
\partial_if_{ij}=0,
\label{em_condition}
\end{equation}
holds, where
$f_{\alpha\beta}\equiv\partial_{\alpha}u_{\beta}-\partial_{\beta}u_{\alpha}$.

\subsection{GT-metrics with flat background}

First, let us consider  a G\"odel-type metric  with flat
background. In the usual cylindrical coordinates
($r,\varphi,z$) the line element for this spacetime  is,
\begin{equation}
ds^2=(dt-\alpha r^2d\varphi)^2-(dr^2+r^2d\varphi^2+dz^2).
\label{gtf}
\end{equation}

 The curve (\ref{CTC}),  $\gamma$,  is timelike
when $g_{\varphi\varphi}=(\alpha^2 r_*^2-1)r_*^2>0$ that leads us to the 
condition,
\begin{equation}
r_*^2>1/\alpha^2.
\label{gtftl}
\end{equation}

For the CTC $\gamma$ we find that the nonzero component of the
four-acceleration satisfies $a^r=r_*(2\alpha^2
r_*^2-1)\dot{\varphi}^2$. The component $a^r$ is identically null when
$r_*^2=1/2\alpha^2$. Therefore the condition for $\gamma$ to be
timelike (\ref{gtftl}) is not satisfied and then this curve can not be
a closed timelike geodesic.

For this CTC the system of perturbation~(\ref{Pertsystem}) can be written as
\begin{eqnarray}
& &\ddot{\xi^0} + k_1 \dot{\xi}^1 = 0, \nonumber \\ 
& &\ddot{\xi^1} + k_2 \dot{\xi}^0 + k_3 \dot{\xi}^2 + k_4 \xi^1 = 0, \nonumber \\ 
& &\ddot{\xi^2} + k_5 \dot{\xi}^1 = 0 \nonumber \\ 
& &\ddot{\xi^3} = 0.
\label{PertSystem4}
\end{eqnarray}
where 
\begin{equation}
k_1=2\Gamma^0_{12}\dot{\varphi},\, k_2=2\Gamma^1_{02}\dot{\varphi},\,
 k_3=\Gamma^1_{22}\dot{\varphi}^2, \,
 k_4=-\Gamma^1_{22}\partial_r(\dot{\varphi}^2),
 \,k_5=2\Gamma^2_{12}\dot{\varphi}.
\end{equation}

The solution of system (\ref{PertSystem4}) is given by:
\begin{equation}
\begin{array}{l}
\xi^0=-k_1(c_3\sin(\omega s+c_4)/\omega+\lambda s)+c_1\,s+c_5,\\ 
\xi^1=c_3\cos(\omega s+c_4)+\lambda, \\
\xi^2=-k_5(c_3\sin(\omega s+c_4)/\omega+\lambda s)+c_2\,s+c_6,\\ 
\xi^3=c_7\,s+c_8,
\end{array}
\end{equation}
where $c_i,\;i=1,\dots,8$ are integration constants,
\begin{eqnarray}
\omega &= &\sqrt{k_4-k_1k_2-k_3k_4}, \\ &=& \Big(\frac{2(2\alpha^6
r_*^6-4\alpha^4 r_*^4+4\alpha^2 r_*^2-1)\dot{\varphi}^2}{r_*^2(\alpha^2
r_*^2-1)^2}\Big)^{1/2},
\end{eqnarray}
and $\lambda = -k_2c_1/\omega^2$. Thus the CTC is linearly stable when
$\alpha\,r_*>0.6$.

\subsection{GT-metrics with conformally flat background}

 Now we shall  studied a G\"odel-type metric with a conformally flat
background, in this case the line element is,
\begin{equation}
ds^2=(dt-\frac{1}{\rho^4}(a+\rho^3 b)\alpha
r^2d\varphi)^2-\frac{1}{\rho^4}(dr^2+r^2d\varphi^2+dz^2).
\label{gtnf}
\end{equation}
where $\rho$ is the radial distance in $R^3$, $\rho=\sqrt{r^2+z^2}$.

From the geodesic equations we find that the nonzero components of
the four-acceleration associated to the curve (\ref{CTC}) are,
\begin{eqnarray}
&& a^r=-\frac{r_*}{\rho_0^2} [\rho_0^2(1-2 \alpha^2 r_*^2 (a+b
\rho_0^3)^2)+r_*^2(r_*^2 \alpha^2 (\rho_0^3 a b-\rho_0^6 b^2+2
a^2)-2)]\dot{\varphi}^2,\nonumber \\ &&
a^z=\frac{z_*}{\rho_0^2}r_*^2(\alpha^2r_*^2(\rho_0^3a\,b-\rho_0^6b^3+2a^2)-2)\dot{\varphi}^2
\label{gtnforce}
\end{eqnarray}
When $z_*=0$ we have $a^z=0$ and we are left with only one nonzero component of the acceleration,
\begin{equation}
a^{r}=\frac{1}{r_*}(\alpha^2(br_*^3+a)(br_*^3-2a)+r_*^2)\dot{\varphi}^2.
\label{gtnforce1}
\end{equation}
We have also $g_{\varphi\varphi}=(\alpha^2(a+br_*^3)^2-r_*^2)/r_*^4$. It is possible
to choose parameters $\alpha$, $a$ and $b$ such that $a^r(r_*)=0$ has
positive roots (see for instance the values presented at the end of
this Sub-Section). Therefore we have CTGs for these values of $r_*$. In this
case, for these CTGs the system~(\ref{Pertsystem}) reduces to
\begin{eqnarray}
&&\ddot{\xi}^0+k_1\dot{\xi}^1=0, \nonumber \\
&&\ddot{\xi}^1+k_2\dot{\xi}^0+k_3\xi^1=0, \nonumber \\
&&\ddot{\xi}^2+k_4\dot{\xi}^1=0, \nonumber \\ 
&&\ddot{\xi}^3=0,
\label{PertSystem5}
\end{eqnarray}
where 
\begin{equation}
k_1=2\Gamma^0_{21}\dot{\varphi},\, k_2=2\Gamma^1_{20}\dot{\varphi},\,
 k_3=\Gamma^1_{22,1}\dot{\varphi}^2, \, k_4=2\Gamma^2_{21}\dot{\varphi}.\nonumber
\end{equation}

The solution of (\ref{PertSystem5})  is
\begin{equation}
\begin{array}{l}
\xi^0=-k_1(c_3\sin(\omega s+c_4)/\omega+\lambda s)+c_1\,s+c_5,\\ 
\xi^1=c_3\cos(\omega s+c_4)+\lambda, \\
\xi^2=-k_4(c_3\sin(\omega s+c_4)/\omega+\lambda s)+c_2\,s+c_6,\\ 
\xi^3=c_7\,s+c_8,
\end{array}
\end{equation}
where $c_i,\;i=1,\dots,8$ are integration constants and 
\begin{equation}
\omega=\Big[\big(1+\frac{a+br_*^3}{r_*^4}(2\alpha^2r_*^2-\alpha^4(br_*^3-2a^2))\big)\dot{\varphi}^2\Big]^{1/2}
\end{equation}

In order to have $\omega^2>0$ it is necessary that the second term
inside of branches be positive or less than one. To do this we
can keep $b$ small. For example: a) choosing $a=b=\alpha=1$, we have
 $r_*=1.138684455$ and 
$g_{\varphi\varphi}=2.876589804$ and $\omega^2=9.459585855\dot{\varphi}^2$, b) for
$a=\alpha=1$ and $b=0.5$, we have  $r_*=1.333767038$,
$g_{\varphi\varphi}=0.9483504120$ and
$\omega^2=5.374106554\dot{\varphi}^2$, and c) for $a=2$, $\alpha=1$ and $b=0.1$,
we have $r_*=2.775627312$, and 
$g_{\varphi\varphi}=0.1587448306$ and $\omega^2=4.44597413\dot{\varphi}^2$. Therefore these three examples represent stable CTGs.

\section{The Soares cosmological model case.} 

This model describes a class of inhomogeneous stationary cosmological
solutions of Einstein-Maxwell equations, with rotating dust and
electromagnetic field \cite{soares}. We are interested in  the subclass of these models
with spacetime  topology   $S^3\times \mathbb R$ 
and  with the dust moving along   closed timelike geodesics.

For the metric 
\begin{equation}
ds^2=A_0^2(dt-2\lambda_1 \cos\theta\,d\varphi)^2-dr^2-B_0^2(d\theta^2+\sin^2\theta\,d\varphi^2),
\end{equation}
where $B_0^2=k\Sigma^2-A_0^2\lambda_1^2$ and $A_0$, $\lambda_1$,
$\Sigma$ and $k$ are constants. Note that by definition the time coordinate  is
a periodic variable \cite{soares}.

The nonspacelike geodesics are
described by the tangent vector field
$\dot{x}^{\alpha}=dx^{\alpha}/ds$, with
\begin{eqnarray}
&&
\dot{t}=k_0+\lambda_1\cos\theta\frac{h_0+k_0A_0^2\lambda_1\cos\theta}{3A_0^2\lambda_1^2\cos^2\theta-B_0^2\sin^2\theta},
\nonumber \\ && \dot{r}=r_0, \nonumber \\ &&
\dot{\theta}=\frac{(h_0+k_0A_0^2\lambda_1\cos\theta)^2+A_0^2k_0^2-1-r_0}{B_0},
\nonumber \\ &&
\dot{\varphi}=\frac{h_0+k_0A_0^2\lambda_1\cos\theta}{3A_0^2\lambda_1^2\cos^2\theta-B_0^2\sin^2\theta},
\nonumber
\end{eqnarray}
where $h_0$, $k_0$ and $r_0$ are arbitrary parameters. Two trivial
cases are given by choosing $\theta=\theta_0=$constant.

{\bf Case1.} Choose $r_0$, $h_0$, $k_0$ such that
\begin{eqnarray}
&& A_0^2k_0^2=1, \nonumber\\
&& r_0=0. \nonumber\\
&& h_0+k_0A_0^2\lambda_1\cos\theta_0=0.
\end{eqnarray}
In this case (\ref{Pertsystem}) can be cast as,
\begin{eqnarray}
&&\ddot{\xi}^0+a\,\dot{\xi}^2=0,\nonumber\\
&&\ddot{\xi}^1=0, \nonumber\\
&&\ddot{\xi}^2+b\,\dot{\xi}^3=0,\nonumber\\
&&\ddot{\xi}^3+c\,\dot{\xi}^2=0,\nonumber
\label{PertSystem3}  
\end{eqnarray}
where $a=-4\lambda_1^2\cos\theta_0A_0^2k_0^2/(B_0^2\sin\theta_0)$,
$b=-2\lambda_1A_0^2k_0^2/(B_0^2\sin\theta_0)$ and
$c=-2\lambda_1k_0^2\sin\theta_0A_0^2/B_0^2$.

The solution of this system is given by
\begin{eqnarray}
&& \xi^0=-a(c_5\,\exp(\omega\,s)/\omega-c_6\,\exp(-\omega\,s)/\omega+bc_4s/\omega^2)+c_1s+c_7;\nonumber \\
&& \xi^1=c_2s+c_3;\nonumber \\
&& \xi^2=c_5\,\exp(\omega\,s)+c_6\,\exp(-\omega\,s)+bc_4/\omega^2;\nonumber \\
&& \xi^3=-c(c_5\,\exp(\omega\,s)/\omega-c_6\,\exp(-\omega\,s)/\omega+bc_4s/\omega^2)+c_4s+c_8.
\end{eqnarray}
where $c_i,\;i=1,\dots,8$ are integration constants, and $\omega=\sqrt{bc}$.
Therefore these CTGs are  not  stable.

{\bf Case2.} Choose $h_0$, $k_0$ such that
\begin{eqnarray}
&& A_0^2k_0^2=1+r_0^2, \nonumber\\
&& h_0+k_0A_0^2\lambda_1\cos\theta_0=0.
\end{eqnarray}
The system of perturbation is the same as before and also the CTGs are not stable.

\section{Discussion}

In summary, we analyzed the linear stability of closed timelike
geodesics in four solutions of Einstein's field equations. It is
possible to find CTGs in different spacetimes, they can be
fulfilled by matter or not and in both cases the CTGs can be
 stable or not. 

In the first case, CTGs outside of an infinite dust cylinder, we found
stable CTGs. In this model the
closed curves are circles in a $(r,\varphi)$-plane $\Pi=\{t=t_*,z=z_*\}$
and all conclusions obtained are independent of the values of the
$t_*$ and $z_*$. In $\Pi$ the CTGs appears in infinitely many regions
(open flat rings) filled by CTCs and these regions are separated by other
regions where the closed curves are spacelike.

In the case of the cloud of cosmic strings there exist CTGs inside the
source but these are not linearly stable. There are CTCs in the
exterior, but no CTG. This is the only case when the matter content of the solution is exotic, i.e., it does not obey the usual energy conditions. We have negative matter density.

Examples of ``cosmological"  solutions with CTCs are   G\"odel-type metric with flat background, as well as,  conformally
flat background. The first has stable CTCs
but no CTGs and for the second it is possible to find values for the parameters
to have a spacetime with  stable CTGs. 

In the ``cosmological" model described
by Soares we  found two examples of not stable CTGs. In  this is a case  the
existence of the CTGs depends upon the nontrivial topology of
spacetime. 

\vspace{0.5cm}
\noindent
{\bf Acknowledgments}
\vspace{0.5cm}

\noindent
V.M.R.  thanks Departamento de Matem\'atica-UFV for giving
the conditions to finish this work which was partially supported  by
PICDT-UFV/CAPES. PSL thanks the partial financial support of FAPESP and
CNPq.


\end{document}